\documentstyle[12pt]{article}
\newcommand{\p}[1]{(\ref{#1})}
\newcommand{\be}{\begin{equation}}
\newcommand{\bea}{\begin{eqnarray}}
\newcommand{\ee}{\end{equation}}
\newcommand{\eea}{\end{eqnarray}}
\newcommand{\cp}{\mbox{$\cal P$}}
\newcommand{\e}{\eta}
\newcommand{\la}[1]{\langle S_{#1}| }
\newcommand{\ra}[1]{|S_{#1}\rangle }

\begin{document}
\topmargin -1cm
\oddsidemargin=0.25cm\evensidemargin=0.25cm
\setcounter{page}0
\renewcommand{\thefootnote}{\fnsymbol{footnote}}
\thispagestyle{empty}
{\hfill  Preprint JINR E2-96-408}\vspace{1.5cm} \\
\begin{center}
{\large\bf  On the BRST approach to the description of a Regge trajectory
}\vspace{0.5cm} \\
A. Pashnev\footnote{E-mail: pashnev@thsun1.jinr.dubna.su}\\
and M. Tsulaia\footnote{E-mail: tsulaia@thsun1.jinr.dubna.su}
\vspace{0.5cm} \\
{\it JINR--Bogoliubov Theoretical Laboratory,         \\
141980 Dubna, Moscow Region, Russia} \vspace{1.5cm} \\
{\bf Abstract}
\end{center}
\vspace{1cm}

The free field theory for Regge trajectory is described in the framework
of the BRST - quantization method. The physical spectrum includes
daugther trajectories along with parent one.
The applicability of the BRST approach to the description of a single
Regge trajectory without its daughter trajectories is discussed.
The simple example illustrates the appropriately modified BRST construction
for the needed second class constraints.

\vspace{0.5cm}
\begin{center}
{\it Submitted to Modern Physics Letters A}
\end{center}

\newpage\renewcommand{\thefootnote}{\arabic{footnote}}
\setcounter{footnote}0\setcounter{equation}0
\section{Introduction}

The main problem in the description of the higher spin particles
is removing of unphysical degrees of freedom from the theory.
As was shown in \cite{FP} the corresponding lagrangian must have
some invariance which generalizes the gauge invariance of the
electromagnetic field not only in the free case but for interacting
fields as well.

On the free level such lagrangians was constructed both for massive
\cite{SH}-\cite{C} and for massless particles \cite{F}-\cite{FF}
of any spin as well as for massless supermultiplets \cite{Cu}.
In some sense the massless case is simpler and, hence, more investigated
than the massive one. Some types of interacting lagrangians
was constructed in the light-cone gauge \cite{BBB} and some progress was
made in the covariant description \cite{BBD} of interactions of
massless higher spin
particles. The hope is that with the help of some Higgs-type effect some
of the interacting particles  acquire nonzero values of the mass.
Up to now the only existing example of higher spin
interacting massive matter is the string theory.

Naturally higher spin particles arise after quantization of classical
extended objects such as string, relativistic oscillator \cite{K}-
\cite{BD},
discrete string \cite{GP} etc. Physically they correspond to exited levels
of the system and belong to Regge trajectories, each including
infinite sequence of states with a spin linearly depending on the
square of the mass. There are infinite number of Regge trajectories
in the string and relativistic oscillator models and only one such
trajectory in the discrete string model due to existing of additional
second-class constraints. Such subdivision of all higher spin particles
on Regge trajectories leads to consideration of these Regge trajectories
as independent objects for which it would be interesting to construct
the lagrangian description.

It seems that one can not construct a consistent interaction of some
finite set of higher spins until all higher spins up to infinity are
included. This naturally leads to consideration of the whole families
of the higher spin particles like Regge trajectory.
 In the infinite limit of the Regge slope
all of the particles have the same value of the mass.
In particular, the case
when all of the particles have zero mass and this Regge trajectory is
vertical {massless tower of particles} is very interesting.

One of the most economic and straightforward method of construction of
lagrangians in gauge theories is the method using BRST-charge of the
corresponding firstly quantized theory. With the help of BRST-charge
the lagrangian of the free field theory of infinite tower of massless
higher spin particles was constructed \cite{OS}. This lagrangian
describes the system which is infinitely degenerated on each spin level
and only with the help of additional constraints we can delete all the
extra states having precisely one particle of each spin . These additional
constraints are of the second class and their inclusion in the BRST-
construction is nontrivial.

The methods of such construction
were discussed in \cite{FS}-\cite{EM}.
With the help of additional variables
one can modify the second class constraints in such a way that they
become commuting, i.e the first class . At the same time the number
of physical degrees of
freedom for both systems of constraints is the same if the number of
additional variables coincides with the number of second class
constraints.

In the second part of the paper we describe the most general system
of constraints. We discuss also the possible truncations and
simplifications of this general systems of constraints.
In the third part of the paper we show
that one of the truncated subsystem of constraints (first- and
second-class)   admits the
construction of BRST-charge along the line of \cite{FS}-\cite{EM}.
As a result the correct free field lagrangian
for Regge trajectory together with its daughter trajectories is
constructed.
The inclusion of additional second class constraints removing the daughter
trajectories is discussed in the fourth part of the article. We
describe the simple modification of the method of \cite{FS}-\cite{EM}
which can be used for the BRST - construction of a single Regge
trajectory.

\section{The systems of constraints}

To describe all higher spins simultaneously it is convenient to
introduce auxiliary Fock space generated by creation and annihilation
operators $a_\mu^+,a_\mu$
with vector Lorentz index $\mu =0,1,2,...D-1$, satisfying the following
commutation relations
\be
[ a_\mu,a_\nu^+ ] =-g_{\mu \nu},\;g_{\mu \nu}=diag(1,-1,-1,...,-1).
\ee
In addition the operators $a_\mu^+,a_\mu$  can have some internal
indices leading to more complicated spectrum of physical states.
For simplicity we consider in this paper $a_\mu^+,a_\mu$
without additional indices.

The general state of the Fock space
\be
|\Phi\rangle =\sum \Phi^{(n)}_{\mu_1\mu_2\cdots\mu_n}(x)a_{\mu_1}^{+}
a_{\mu_2}^+\cdots a_{\mu_n}^+ |0\rangle
\ee
depends on space-time coordinates $x_\mu$ and its components
$\Phi^{(n)}_{\mu_1\mu_2\cdots\mu_n}(x)$ are tensor fields of rank $n$
in the space-time of arbitrary dimension $D$.
The norm of states in this Fock space is not positively definite due to
the minus sign in the commutation relation (2.1) for time components of
creation and annihilation operators. It means that physical states must
satisfy some constraints to have positive norm. These constraints
arise naturally in the considerations of classical composite systems
\cite{BD}-\cite{GP}.
The corresponding quantum operators
\bea \label{GEN}
L_0 &=&-p^2-\alpha'a_\mu^+a_\mu,\;\;\;
L_1 = pa,\;\;\; L_{-1}=pa^+ =L_1^+,\\ \label{l2}
L_2 &=& \frac{1}{2}aa,\;\;\; L_{-2}=\frac{1}{2}a^+a^+=L_2^+
\eea
form the algebra
\bea
[L_0\;,\;L_{\pm 1}]=\mp\alpha'L_{\pm 1} & [L_0\;,\;L_{\pm 2}]= &
\mp 2\alpha'L_{\pm 2}\\
\eea
\bea
[L_{1}\;,\;L_{-2}]=-L_{-1} & [L_{-1}\;,\;L_2]= & L_{1}\\
\eea
\bea
[L_1\;,\;L_{-1}]=-p^2 & [L_2\;.\;L_{-2}] =&-a_\mu^+a_\mu+\frac{D}{2}\equiv
G_0
\eea
The operators $L_0, L_1, L_2$ correspond to the mass shell,
transversality and tracelessness conditions on the wavefunctions.
The operators $L_1 , L_{-1}$ and  $L_2 , L_{-2}$ are of second class
and in general this system of constraints describes single Regge
trajectory \cite{BD}-\cite{P}.

There exist some different possibilities in consideration of
general system of constraints \p{GEN}-\p{l2}.
 The truncated system $L_0 , L_{\pm 1}$
describes Regge trajectory together with its
daughter trajectories. The operators $L_{\pm 1}$ in this system
are of second class as before. We describe the BRST - quantization
of this system in the third part of the article.

The limit $\alpha'=0$ of the system \p{GEN}
corresponds to the massless infinite tower
of spins with a single state  at each value of the spin.
In this case only the operators $L_{\pm 2}$ are of second class.
The BRST - quantization of this system as well as general system
\p{GEN}-\p{l2} will be given elsewhere.

The simplest system of first-class constraints
 \begin{equation}\label{OS}
\tilde{L}_0=-p^2\;,\;L_{\pm 1}
\end{equation}
corresponds to the massless tower of spins infinitely degenerated
at each value of spin. The BRST - construction for this system of
constraints was described in \cite{OS}.

\setcounter{equation}0\section{BRST-quantization of the Regge trajectory}

In this section we consider the system with constraints
\begin{equation}
L_0 =-p^2-\alpha'a_\mu^+a_\mu+\alpha_0,\;\;\;
L_1 = pa,\;\;\; L_1^+=pa^+,
\end{equation}
where parameter $\alpha_0$
 plays the role of intercept
for Regge trajectory.
 In some sense this system is
intermediate  between the systems in \cite{OS} and \cite{P} because
it describes the Regge trajectory together with all daughter trajectories
(due to the absence of the constraints $L_{\pm 2}$).
The commutation relation $[L_1\;,\;L_{-1}]=-p^2 $ means that
$L_{\pm 1}$ are the second class   constraints .
Following the prescription of \cite{FS} - \cite{EM} we
introduce  operators $b$ and $b^+$ with the commutation relations
$[b,b^+]=1$ and modify  the constraints to the following expressions:
\begin{eqnarray}            \label{AAA}
\tilde L_0& = &L_0 + \alpha'b^+b+\alpha',\\
\tilde L_{-1}& = &L_{- 1} + \sqrt{p^2}b^+,\\
\tilde L_{1}& = &L_{1} + \sqrt{p^2}b,
\end{eqnarray}

\begin{equation}\label{A1}
[\tilde L_1 , \tilde L_{-1}]=0, \quad [\tilde L_0,\tilde L_1] =
- \alpha'\tilde L_1,   \quad [\tilde L_0,\tilde L_{- 1}] =
\alpha'\tilde L_{-1}.
\end{equation}

All of the modified constraints are of first class and BRST - charge
construction is straightforward.
Firstly we
introduce additional set of anticommuting variables
$\e_0,\e_1,\e_1^+$ having ghost number one and corresponding momenta
$\cp_0,\cp_1^+,\cp_1$ with commutation relations:
\begin{equation}
\{\e_0,\cp_0\}=\{\e_1,\cp_1^+\}=\{\e_1^+,\cp_1\}=1.
\end{equation}
The nilpotent BRST - charge has the following form:
\begin{equation}
Q = \e^+_1 \tilde L_1 + \e_1 \tilde L_{-1} + \e_0 \tilde L_0 +
\alpha'\e_0 \e^+_1 \cp_1 - \alpha'\e_0 \e_1 \cp^+_1
\end{equation}
Consider the total Fock space generated by creation operators
$a_\mu^+,b^+,\e_1^+,\cp_1^+$. In addition each vector of the Fock
space depends linearly on the real grassmann variable $\e_0$
($\cp_0$ considered as corresponding derivative
$\cp_0=\partial / \partial\e_0$)
\begin{equation}
|\chi\rangle =  |\chi_1\rangle +\e_0 |\chi_2\rangle.
\end{equation}
Ghost numbers of $|\chi_1\rangle$ and $|\chi_2\rangle$ are different if
the state $|\chi\rangle$ has some definite one.

The BRST - invariant lagrangian in such Fock space can be written as
\begin{equation} \label{L11}
L=\int d \e_0 \langle\chi|Q|\chi\rangle.
\end{equation}
To be physical, lagrangian $L$ must have zero ghost number. It means
that vectors $|\chi\rangle$ and $|\chi_1\rangle$ must have zero ghost
numbers as well. In this case the ghost number of $|\chi_2\rangle$
is minus one. The most general expressions for such vectors are
\begin{eqnarray}
|\chi_1\rangle&=&\ra{1}+\e_1^+\cp_1^+\ra{2},\\
|\chi_2\rangle&=&\cp_1^+\ra{3},
\end{eqnarray}
with vectors $\ra{i}$ having ghost number zero and depending only
on bosonic creation operators $a_\mu^+,b^+$
\begin{equation}\label{RA}
\ra{i}=\sum \phi^{n}_{\mu_1,\mu_2,...\mu_n}(x)
a_{\mu_1}^+ a_{\mu_2}^+ ...a_{\mu_n}^+(b^+)^{n}|0\rangle.
\end{equation}

Integration over the $\e_0$ leads to the following lagrangian in terms
of $\ra{i}$
\begin{eqnarray}\label{LR}
L&=&\la{1}\tilde L_0 \ra{1} - \la{2}\tilde L_0 \ra{2} -
\alpha'\la{1}\ra{1} - \alpha'\la{2}\ra{2}-\\
&&-\la{1}\tilde L^+_{1}\ra{3} - \la{3}\tilde L_1 \ra{1} +
\la{2}\tilde L_1 \ra{3} + \la{3} \tilde L^+_{1} \ra{2}
\end{eqnarray}
For comparison with the massless case we write down the lagrangian
of \cite{OS} in the same terms
\begin{eqnarray}\label{LT}
L&=&-\la{1}p^2 \ra{1} + \la{2}p^2 \ra{2} + \la{3}\ra{3}- \\
&&-\la{1} L^+_{1}\ra{3} - \la{3} L_1 \ra{1} +
\la{2} L_1 \ra{3} + \la{3} L^+_{1} \ra{2}
\end{eqnarray}

The nilpotency of the BRST - charge leads to the invariance of the
lagrangian \p{L11} under the following transformations
\begin{equation}\label{TR}
\delta |\chi \rangle = Q|\Lambda \rangle.
\end{equation}
The parameter of transformation must have ghost number $-1$ and
can be written as  $|\Lambda \rangle = \cp^+_1|\lambda\rangle ,$
where $|\lambda\rangle$ belong to the Fock space generated by
$a_\mu^+ , b^+$ and depends from the space - time coordinates.
On the component level such invariance leads
to the invariance of the lagrangian \p{LR} under the following
transformations
\begin{eqnarray}\label{To1}
\delta\ra{1}&=&\tilde{L}_1^+|\lambda\rangle,\\ \label{To2}
\delta\ra{2}&=&\tilde{L}_1|\lambda\rangle,\\ \label{To3}
\delta\ra{3}&=&\tilde{L}_0|\lambda\rangle.
\end{eqnarray}

One can show that using together \p{To1} and equations of motion
for the fields $\ra{i}$
\begin{eqnarray}
(\tilde L_0 -\alpha')\ra{1}&=&\tilde L^+_1 \ra{3},\\
(\tilde L_0 +\alpha')\ra{2}&=&\tilde L_1 \ra{3}, \\
\tilde L_1 \ra{1}&=&\tilde L^+_1 \ra{2},
\end{eqnarray}
one can eliminate the fields $\ra{2}$ and $\ra{3}$. Firstly we
solve the equation
\begin{equation}\label{S3}
\ra{3}+\tilde L_0|\lambda \rangle =0
\end{equation}
using decompositions
\begin{equation}
|S_i\rangle =\sum (b^+)^n|S_{in}\rangle,
|\lambda\rangle =\sum (b^+)^n|\lambda_{n}\rangle.
\end{equation}
The equation \p{S3} does not fix parameter $|\lambda\rangle$ completely.
There will be residual invariance with parameter
$|{\lambda}^{\prime}\rangle$ under the condition
$\tilde L_0|{\lambda}^{\prime}\rangle=0$.
After the elimination of the field $|S_2\rangle$
with the help of the equation $
\ra{2}+\tilde L_1|{\lambda}^{\prime}\rangle=0
$
the new parameter
$|{\lambda}^{\prime\prime}\rangle$ will satisfy two conditions
$\tilde L_0|{\lambda}^{\prime\prime}\rangle=
\tilde L_1|{\lambda}^{\prime\prime}\rangle=0$.
With the help of this parameter all fields $|S_{1n}\rangle$,
except zero mode $|S_{10}\rangle$ can be eliminated.
The residual gauge invariance $\delta|S_{10}\rangle=
L_1^+|\lambda^{\prime\prime\prime}\rangle$ allows to impose constraint
\begin{equation}\label{CON}
L_1 |S_{10}\rangle=0,
\end{equation}
 which together with the equation of motion
\begin{equation}\label{EQU}
L_0 |S_{10}\rangle=0
\end{equation}
 describe the spectrum of Regge trajectory. The constraint \p{CON}
kills unphysical degrees of freedom and equation \p{EQU} fixes
linear dependence between spin and square of mass. The absence
of the condition
\begin{equation}\label{CON2}
L_2 |S_{10}\rangle=0
\end{equation}
means that the wavefunctions $\phi_{\mu_1,\mu_2,...\mu_n} $
belong to the reducible representations of the $D$- dimensional
Lorentz group. In turn it means that the daughter Regge
trajectories belong to the spectrum as well.

\setcounter{equation}0\section{The simple example}
To describe the Regge trajectory without its daughter trajectories
we must take into account additional constraint \p{CON2} on the
wavefunctions. It means that the second class constraints
$L_{\pm 2}$ must be included in the BRST charge. Following the line
of \cite{FS} - \cite{EM} we can transform these constraints into
commuting ones by introducing additional freedoms like $b, b^+$.
This procedure is rather simple for the classical
case when Poisson brackets are used instead of commutators.
It leads to the finite system of differentional equations which can be
solved without troubles.
In the quantum case the corresponding system of equations is infinite
due to accounting of repeated commutators.

In this section we describe the modification of the procedure
of \cite{FS} - \cite{EM} which works well in the case of two
second class constraints
$L_2$ and $L_{-2}$ \p{l2}.  We modify this system of constraints
by introduction of {\it two} additional operators $b_1, b_2$ together with
their conjugates $b_1^+, b_2^+$: $[b_i, b_k^+]=\delta_{ik}$. New
constraints are
\begin{eqnarray}
\tilde{L}_2&=&L_{2}+b_1^+b_2, \\
\tilde{L}_{-2}&=&L_{-2}+b_2^+b_1.
\end{eqnarray}
Together with new operator $\tilde{G}_0=G_0+b_2^+b_2-b_1^+b_1$ they
form an $SU(2)$ algebra
\begin{eqnarray}
[ \tilde{L}_2 \; , \; \tilde{L}_{-2} ] &=& \tilde{G}_0, \\
\;[ \tilde{G}_0 \; , \; \tilde{L}_{\pm 2} ] &=&\mp2\tilde{L}_{\pm 2}.
\end{eqnarray}
and can be considered as first class constraints. The counting of
physical degrees of freedom shows equal number for both systems
of constraints. Indeed, we have introduced four additional variables
$b_i, b_k^+$, but instead of two second class constraints each killing
one degree of freedom we get three first class constraints which
kill together six degrees of freedom.

To illustrate the BRST - approach to this simple system we
introduce additional set of anticommuting variables
$\e_0,\e_2,\e_2^+$ having ghost number one and corresponding momenta
$\cp_0,\cp_2^+,\cp_2$ with commutation relations:
\begin{equation}
\{\e_0,\cp_0\}=\{\e_2,\cp_2^+\}=\{\e_2^+,\cp_2\}=1.
\end{equation}
The standard prescription
gives the following nilpotent BRST - charge
\begin{equation}
Q=\e_2^+\tilde{L}_2+\e_2\tilde{L}_2^++\e_0\tilde{G}_0+
  \e_2\e_2^+\cp_0+2\e_0\e_2^+\cp_2-2\e_0\e_2\cp_2^+.
\end{equation}

Consider the total Fock space generated by creation operators
$a_\mu^+,b_i^+,\e_2^+,\cp_2^+$. In addition each vector of the Fock
space depends linearly on the real grassmann variable $\e_0$
($\cp_0$ considered as corresponding derivative
$\cp_0=\partial / \partial\e_0$)
\begin{equation}
|\chi\rangle =  |\chi_1\rangle +\e_0 |\chi_2\rangle.
\end{equation}
Ghost numbers of $|\chi_1\rangle$ and $|\chi_2\rangle$ are different if
the state $|\chi\rangle$ have some definite one.

The BRST - invariant lagrangian in such Fock space can be written as
\begin{equation} \label{L1}
L=\int d \e_0 \langle\chi|Q|\chi\rangle.
\end{equation}
To be physical, lagrangian $L$ must have zero ghost number. It means
that vectors $|\chi\rangle$ and $|\chi_1\rangle$ must have zero ghost
number as well. In this case the ghost number of $|\chi_2\rangle$
is minus one. The most general expressions for such vectors are
\begin{eqnarray}
|\chi_1\rangle&=&\ra{1}+\e_2^+\cp_2^+\ra{2},\\
|\chi_2\rangle&=&\cp_2^+\ra{3},
\end{eqnarray}
with vectors $\ra{i}$ having ghost number zero and depending only
on bosonic creation operators $a_\mu^+,b_i^+$
\begin{equation}
\ra{i}=\sum \phi^{n_1,n_2}_{\mu_1,\mu_2,...\mu_n}
a_{\mu_1}^+ a_{\mu_2}^+ ...a_{\mu_n}^+(b_1^+)^{n_1}(b_2)^{n_2}|0\rangle.
\end{equation}
In general the vawefunction $\phi^{n_1,n_2}_{\mu_1,\mu_2,...\mu_n} $
can depend on other physical variables of the theory such as
space - time coordinates etc.

Integration over the $\e_0$ leads to the following lagrangian in terms
of $\ra{i}$
\begin{eqnarray}             \label{L2}
L&=&\la{1}\tilde{G}_0\ra{1}-2\la{1}\ra{1}
 -\la{2}\tilde{G}_0\ra{2}-2\la{2}\ra{2} +\la{3}\ra{3}-\\
&&-\la{1}\tilde{L}_2^+\ra{3}-\la{3}\tilde{L}_2\ra{1}+
\la{2}\tilde{L}_2\ra{3}+\la{3}\tilde{L}_2^+\ra{2}.\nonumber
\end{eqnarray}

Owing to the nilpotency of the BRST -- charge - $Q^2=0$, the lagrangian
\p{L1} is invariant under the transformation
\begin{equation}
\delta |\chi\rangle =Q |\Lambda\rangle
\end{equation}
with $|\Lambda\rangle=\cp_2^+|\lambda\rangle$. In turn it means the
invariance of the lagrangian \p{L2} under the following transformation
\begin{eqnarray}\label{Tr1}
\delta\ra{1}&=&\tilde{L}_2^+|\lambda\rangle,\\ \label{Tr2}
\delta\ra{2}&=&\tilde{L}_2|\lambda\rangle,\\ \label{Tr3}
\delta\ra{3}&=&\tilde{G}_0|\lambda\rangle.
\end{eqnarray}

The field $\ra{3}$ is auxiliary. Using its equation of motion
\begin{equation}
\ra{3}=\tilde{L}_2\ra{1}-\tilde{L}_2^+\ra{2},
\end{equation}
the lagrangian \p{L2} can be rewritten in terms of two fields
$\ra{1}$ and $\ra{2}$
\begin{eqnarray}
L&=&\la{1}(\tilde{G}_0-2-\tilde{L}_2^+\tilde{L}_2)\ra{1}-
\la{2}(\tilde{G}_0+2+\tilde{L}_2\tilde{L}_2^+)\ra{2}+\\
&&+\la{1}\tilde{L}_2^+\tilde{L}_2^+\ra{2}+
\la{2}\tilde{L}_2\tilde{L}_2\ra{1}\nonumber
\end{eqnarray}
with the following equations of motion:
\begin{eqnarray}
&&(\tilde{G}_0-2-\tilde{L}_2^+\tilde{L}_2)\ra{1}+
\tilde{L}_2^+\tilde{L}_2^+\ra{2}=0,\\
&&(\tilde{G}_0+2+\tilde{L}_2\tilde{L}_2^+)\ra{2}-
\tilde{L}_2\tilde{L}_2\ra{1}=0.
\end{eqnarray}

Following the line of preceding section one can show that the
gauge freedom \p{Tr1}-\p{Tr3} is sufficient to kill the
$b_2^+$ dependence in $\ra{1}$ and get the following
conditions on the reduced field $|\tilde{S}_1\rangle$:
\begin{equation}\label{LAST}
\frac{1}{2}aa|\tilde{S}_1\rangle=0,\;
(-a^+a + \frac{D}{2}-b^+_1 b_1)|\tilde{S}_1\rangle=0.
\end{equation}
The first of the conditions \p{LAST} means tracelessness of the
wavefunctions $\phi^{n_1,0}_{\mu_1,\mu_2,...\mu_n} $. The second one
connects $n$ and $n_1$: $n_1=n+\frac{D}{2}$ killing effectively the
$b_1^+$ dependence in $\ra{1}$.  So, the system contains no
additional degrees of freedom and describes traceless wavefunctions.

\setcounter{equation}0\section{Conclusions}
In this paper we have applied the BRST approach to the description
of the free Regge trajectory. We have described also the simple
modification of the conversion procedure for the second class
constraints which can be used for elimination of daughter
trajectories from the physical spectrum. The corresponding
modification of the BRST approach to the single Regge
trajectory will be given elsewhere.
\vspace{1cm}

\noindent {\bf Acknowledgments.}
This investigation has been supported in part by the
Russian Foundation of Fundamental Research,
grants 96-02-17634 and 96-02-18126, joint grant RFFR-DFG 96-02-00186G,
and INTAS, grant 94-2317 and grant of the
Dutch NWO organization.

\end{document}